\documentclass[a4paper]{article}

\usepackage{icrc2013}
\usepackage[english]{babel}

\newcommand\beq{\begin{equation}}
\newcommand\beql[1]{\begin{equation} \label{#1}}
\newcommand\eeq{\end{equation}}
\newcommand\ben{\begin{eqnarray}}
\newcommand\een{\end{eqnarray}}
\newcommand\bea{\begin{array}}
\newcommand\beal[1]{\begin{array} \label{#1}}
\newcommand\eea{\end{array}}
\newcommand\bem{\begin{displaymath}}
\newcommand\eem{\end{displaymath}}

\newcommand\eqa[1]{Eq.(\ref{#1})}
\newcommand\eqb[1]{Eqs.(\ref{#1})}
\newcommand\eqc[1]{(\ref{#1})}

\newcommand\fig[1]{Fig.\ref{#1}}
\newcommand\figs[1]{Figs.\ref{#1}}
\newcommand\figg[1]{\ref{#1}}

\newcommand\app[1]{Appendix \ref{#1}}

\newcommand\qqa{\quad}
\newcommand\qqb{\qquad}

\newcommand\Ssum[2]{\sum \limits_{#1}^{#2}}

\newcommand\dd[2]{\frac{{\rm d} #1}{{\rm d} #2}}

\newcommand\eV{{\rm eV}}
\newcommand\MeV{{\rm MeV}}
\newcommand\TeV{{\rm TeV}}
\newcommand\EeV{{\rm EeV}}

\newcommand\Xmax{X_{\rm max}}
\newcommand\AXmax{\langle X_{\rm max} \rangle}
\newcommand\SXmax{\sigma^{2}_{\rm max}}
\newcommand\sxmax{\sigma_{\rm max}}
\newcommand\ssxmax{\sigma (X_{\rm max})}
\newcommand\Qmax{Q_{\rm max}}

\newcommand\AXmaxA{\langle X_{\rm max} \mid A \rangle}
\newcommand\AAXmaxA{\langle \langle X_{\rm max} \mid A \rangle \rangle}
\newcommand\AXmaxAp{\langle X_{\rm max} \mid A=1 \rangle}

\newcommand\Axmax{\langle x_{\rm max} \rangle}

\newcommand\SXmaxA{\sigma^{2} (X_{\rm max} \mid A)}
\newcommand\SSXmax{\sigma^{2} (X_{\rm max})}
\newcommand\SXmaxAA{\sigma^{2} (\langle X_{\rm max} \mid A \rangle)}
\newcommand\ASXmaxA{\langle \sigma^{2} (X_{\rm max} \mid A) \rangle}

\newcommand\sfr{\sigma_{\rm fr}^{2}}
\newcommand\ssh{\sigma_{\rm sh}^{2}}
\newcommand\saa{\sigma_{\ln {\rm A}}^{2}}
\newcommand\sfrA{\langle \sigma_{\rm fr}^{2} \rangle}
\newcommand\sshA{\langle \sigma_{\rm sh}^{2} \rangle}
\newcommand\sfro{\sigma_{\rm fr,0}^{2}}
\newcommand\ssho{\sigma_{\rm sh,0}^{2}}
\newcommand\sfroo{\sigma_{\rm fr,0}}
\newcommand\sshoo{\sigma_{\rm sh,0}}

\newcommand\pa{p_{\rm A}}
\newcommand\lnAA{\langle \ln A \rangle}
\newcommand\lnAtA{\langle \ln^{2} A \rangle}
\newcommand\lnAs{\sigma_{\rm \ln A}^{2}}

\newcommand\gIcmS{{\rm gcm^{-2}}}
\newcommand\mb{{\rm mb}}

\newcommand\Log{{\rm Log}}
\newcommand\ovl[1]{\overline{#1}}

\newcommand\smu{\sigma_{\rm M}^{2}}
\newcommand\ska{\sigma_{\rm \kappa}^{2}}
\newcommand\smuo{\sigma_{\rm M,0}^{2}}
\newcommand\skao{\sigma_{\rm \kappa,0}^{2}}

\title{Testing chemical composition of highest energy comic rays}

\shorttitle{Decomposition of the mass composition}

\authors{
Dalibor Nosek$^{1}$,
Jakub Vicha$^{2}$,
Jana Noskova$^{3}$,
Jan Ebr$^{2}$.
}

\afiliations{
$^1$ Institute of Particle and Nuclear Physics, 
Faculty of Mathematics and Physics, Charles University, 
Prague, Czech Republic \\
$^2$ Institute of Physics, Academy of Sciences of the Czech Republic, 
Prague, Czech Republic \\
$^3$ Department of Mathematics, Faculty of Civil Engineering, 
Czech Technical University, Prague, Czech Republic
}

\email{nosek@ipnp.troja.mff.cuni.cz}

\abstract{
We study basic characteristics of distributions of the depths
of shower maximum in air showers caused by cosmic rays with
the highest energies. 
The consistency between their average values and widths, and 
their energy dependences are discussed within a simple 
phenomenological model of shower development independently 
of assumptions about detailed features of high--energy interactions. 
It is shown that reliable information on primary species can be 
derived within a partition method. 
We present examples demonstrating implications for
the changes in mass composition of primary cosmic rays.
}


\begin{document}

\maketitle

\section{Introduction}
\label{Sec01}

Knowledge of the mass distribution of cosmic rays (CR) and of its 
energy evolution can provide useful information about CR acceleration 
mechanisms and propagation through the galactic and extragalactic space. 
Measurements and subsequent analysis of the mass composition 
of ultra--high energy cosmic ray (UHECR) primaries are of particular 
importance.
Corresponding observables can help to understand their typical
spectral features, 
the ankle at about $4~\EeV$ and the steep flux suppression 
at energies above $30~\EeV$.
In addition, their knowledge makes searches for the CR sources 
much easier.

In seeking for the mass of UHECR particles, the development 
of extensive air showers (EAS) of secondary particles created 
in the Earth atmosphere is usually examined. 
The mean penetration depth in the atmosphere at which 
the shower of secondary particles reaches its maximum number, 
$\AXmax$, and $\sxmax = \ssxmax$, the square root of its 
variance, are widely used.
Recent results presented by the Auger collaboration 
indicate a transition from lighter to heavier primaries 
at the ankle region~\cite{Aug01,Aug02}.
The HiRes collaboration achieved a different conclusion.
Its analysis based on the truncated fluctuation widths speaks 
in favor of very light primaries at the highest energies~\cite{Hir01}.

Based on widely accepted empirical characteristics of the energy 
evolution of the mean depth of shower maximum and its variance, 
we present a method in which reasonable inferences about 
the partition of the primary CR mass are naturally achieved. 
Utilizing a generalized Heitler model~\cite{Mat01,Ung01}, 
two illustrative examples are presented.
We make use of the recently measured value of the $p$--air 
cross section~\cite{Aug03} and try intentionally to account for 
the details of EAS development independently of assumptions
about detailed features of hadronic interactions.

\section{Air shower model}
\label{Sec02}

Let us assume that a CR shower maximum $\Xmax$ is measured 
when a UHECR particle with a mass $A$ hits the upper part 
of the Earth atmosphere.
In the following we will treat these two quantities as dependent 
random variables, $\Xmax = \Xmax(A)$.  
Adopting superposition assumptions~\cite{Ung01}, the mean depth of 
shower maximum provided air showers are initiated by primaries of 
the mass $A$ depends on the shower energy $E$ as~\cite{Mat01,Ung01}
\beql{E01}
\AXmaxA = C + D~\Log \left( \frac{E}{E_{0} A} \right).
\eeq
Here, $D = \dd{\Axmax}{\Log E}$ is the proton elongation 
rate~\cite{Ung01}, where $\Axmax = \AXmaxAp$
is the proton mean depth of shower maximum, and
$C = \Axmax(E_{0})$ is a constant proton mean depth of maximum 
at a reference energy of $E_{0}$. 
In the same line, the conditional variance of the depth of maximum is
\beql{E02}
\SXmaxA = \sfr + \ssh,
\eeq
where $\sfr = \sfr(A,E)$ is the variance of the depth 
where the first interaction of the CR primary takes place and 
$\ssh = \ssh(A,E)$ assigns the variance of the 
depth of shower maximum associated with its subsequent 
development~\cite{Ung01}.
Then, the total mean and total variance of the depth of shower 
maximum at a given energy $E$ that are to be confronted with 
measurements, are respectively
\beql{E03}
\AXmax = \AAXmaxA = \Axmax - d \lnAA,
\eeq
and
\beql{E04}
\SXmax = \SSXmax = \sfrA + \sshA + d^{2} \saa,
\eeq
where $D = d \ln 10$ was inserted and the law of total variance was 
used, i.e.$\SXmax = \ASXmaxA + \SXmaxAA$, see e.g. Ref.~\cite{Rao01}. 
Except for $\Axmax$, the other mean values written on the right 
hand sides in~\eqb{E03} and~\eqc{E04} are calculated over mass numbers 
of primary CR particles.
\begin{figure}[t!]
\vspace{-2cm}
\centering
\includegraphics[width=0.48\textwidth]{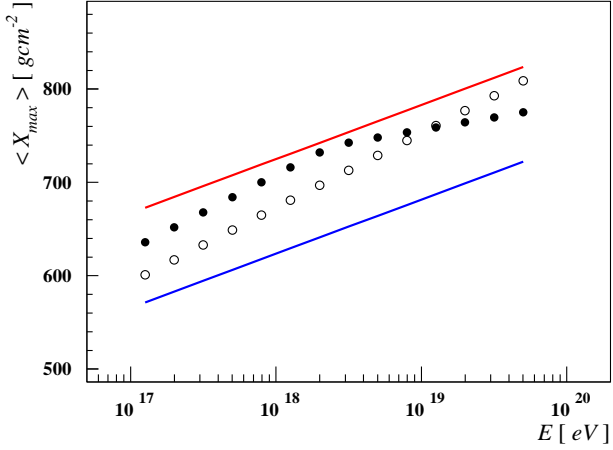}
\caption{
Two hypothetical examples of mean values of $\AXmax$ are shown 
as functions of energy.
Black empty points are for the constant elongation rate.
Black full points show the elongation rate with a break 
as indicated by a magenta arrow.  
MC predictions of $\AXmax$ for proton and iron primaries are 
illustrated by red and blue line, respectively.
}
\label{F01}
\end{figure}

\section{Partition problem}
\label{Sec03}

To examine the mass composition we utilize the partition method
described briefly in~\app{App01}.
To this end, we use two $A$--dependent constrains, 
$F_{1}(A) = d \ln A$ and 
$F_{2}(A) = d^{2} \ln^{2} A + \sfr + \ssh$, respectively.
Their average values are given by the available 
experimental information contained in the $\Xmax$ measurements.
They are directly connected to the total sample mean, $\AXmax$, 
and to the total sample variance of $\Xmax$, $\SXmax$, measured 
at given energy. 
The aforementioned constrains are written as
\beql{E05}
\langle F_{1} \rangle = 
d \lnAA = \Qmax,
\eeq
\beql{E06}
\langle F_{2} \rangle = 
d^{2} \lnAtA + \sfrA + \sshA = \SXmax  + \Qmax^{2},
\eeq
where $\Qmax = \Axmax - \AXmax$.

In the partition method, the probability distribution of the mass 
number is dictated by the maximum--entropy principal as described 
in~\app{App01}. 
Knowing the total mean and variance at given energy, $\AXmax$ and $\SXmax$,
the form of this distribution is given by~\eqa{A03} with two Lagrangian
multipliers deduced numerically in such a way that the two constrains
written in~\eqb{E05} and~\eqc{E06} are satisfied. 

In this study, the proton mean depth of shower maximum 
$C = \Axmax(E_{0})$ at a reference energy of $E_{0}$ and the energy 
independent proton elongation rate $D = d\ln 10$ are only two 
free parameters.

The $A$--dependence of the depth of shower maximum 
is given by the Heitler conjecture, see~\eqa{E01}. 
For other mass dependent terms we use simple phenomenological
arguments described in~\app{App02}. 
The variance of the depth of the first interaction, 
$\sfr = \sfr(A,E)$, is deduced from the measured $p$--air cross
section and its extrapolated energy dependence~\cite{Aug03}.
The variance of the depth of shower maximum connected with the shower 
development, $\ssh = \ssh(A,E)$, is inferred from basic 
characteristics of underlying interaction processes.
Let us stress that other parametrizations of the EAS
$A$--dependent terms, different from that ones introduced 
in~\app{App02}, can be adopted in our treatment.
\begin{figure}[t!]
\vspace{-2cm}
\centering
\includegraphics[width=0.48\textwidth]{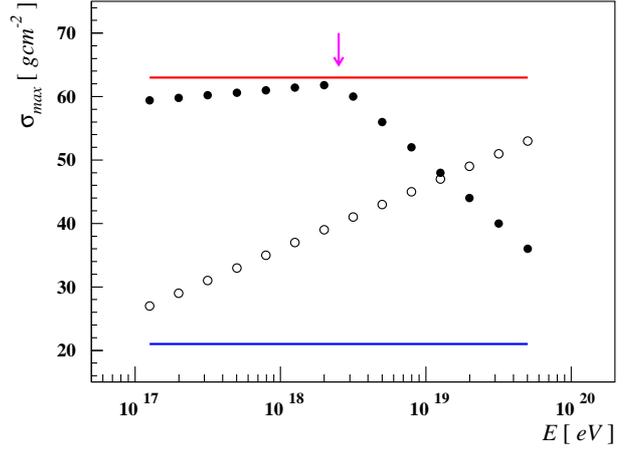}
\caption{
Square roots of variances $\sxmax$ used in two illustrative example 
are shown as functions of energy.
The constant elongation rate (black empty points) and the elongation 
rate with a break (black full points) are shown, see also~\fig{F01}. 
Red and blue lines illustrate MC predictions of 
$\sxmax$ for proton and iron primaries, respectively.
}
\label{F02}
\end{figure}

\section{Illustrative examples}
\label{Sec04}

We successfully applied the maximum--entropy method for a number of 
artificially chosen examples with average shower characteristics 
resembling their measured energy evolution.
Within the partition method, we decomposed these observables into 
different sets of primary masses assuming different parametrization 
of the mean depth of shower maximum and its variance.
In the following, we present results of two of these hypothetical 
examples. 

In the first example, we used the mean depth of shower maximum 
with a constant elongation rate and a logarithmically increasing 
square root of the depth variance with energy.
These shower statistics, displayed in~\figs{F01} and~\figg{F02}
as black empty points, were parametrized by 
\beql{E07}
\frac{\AXmax(E) - X_{0}}{D_{0}} =
\frac{\sxmax(E) - \sigma_{0}}{s_{0}} = 
\Log \left( \frac{E}{E_{0}} \right),
\eeq
where $X_{0} = 673~\gIcmS$ and $\sigma_{0} = 36~\gIcmS$ are shower 
statistics at a reference energy of $E_{0} = 1~\EeV$, and parameters 
$D_{0} = 80~\gIcmS$ and $s_{0} = 10~\gIcmS$.
An energy interval $\Log(E/\eV) \in \langle 17.1, 19.7 \rangle$
with $14$ equidistant values ($\Delta \Log(E/\EeV) = 0.2$) was
assumed.

In the following calculations, we tried to decompose the mass 
composition represented by the shower statistics, $\AXmax(E)$ and
$\sxmax(E)$, into four pieces corresponding to primary species 
generating underlying CR showers.
Namely, we assumed proton primaries ($A=1$), and helium ($A=4$), 
nitrogen ($A=14$) and iron ($A=56$) nuclei.
In the first step, we solved the partition problem numerically 
treating the two unknown quantities, $C$ and $D$
introduced in~\eqa{E01}, as free parameters. 
This way, we obtained a two--dimensional domain where maximum--entropy 
solutions exist.
In the second step, we performed the partition analysis with 
parameters $C = (730-740)~\gIcmS$ and $D = (56-60)~\gIcmS$ that 
provided us the best solutions of the partition problem.

\begin{figure}[t!]
\vspace{-2cm}
\centering
\includegraphics[width=0.48\textwidth]{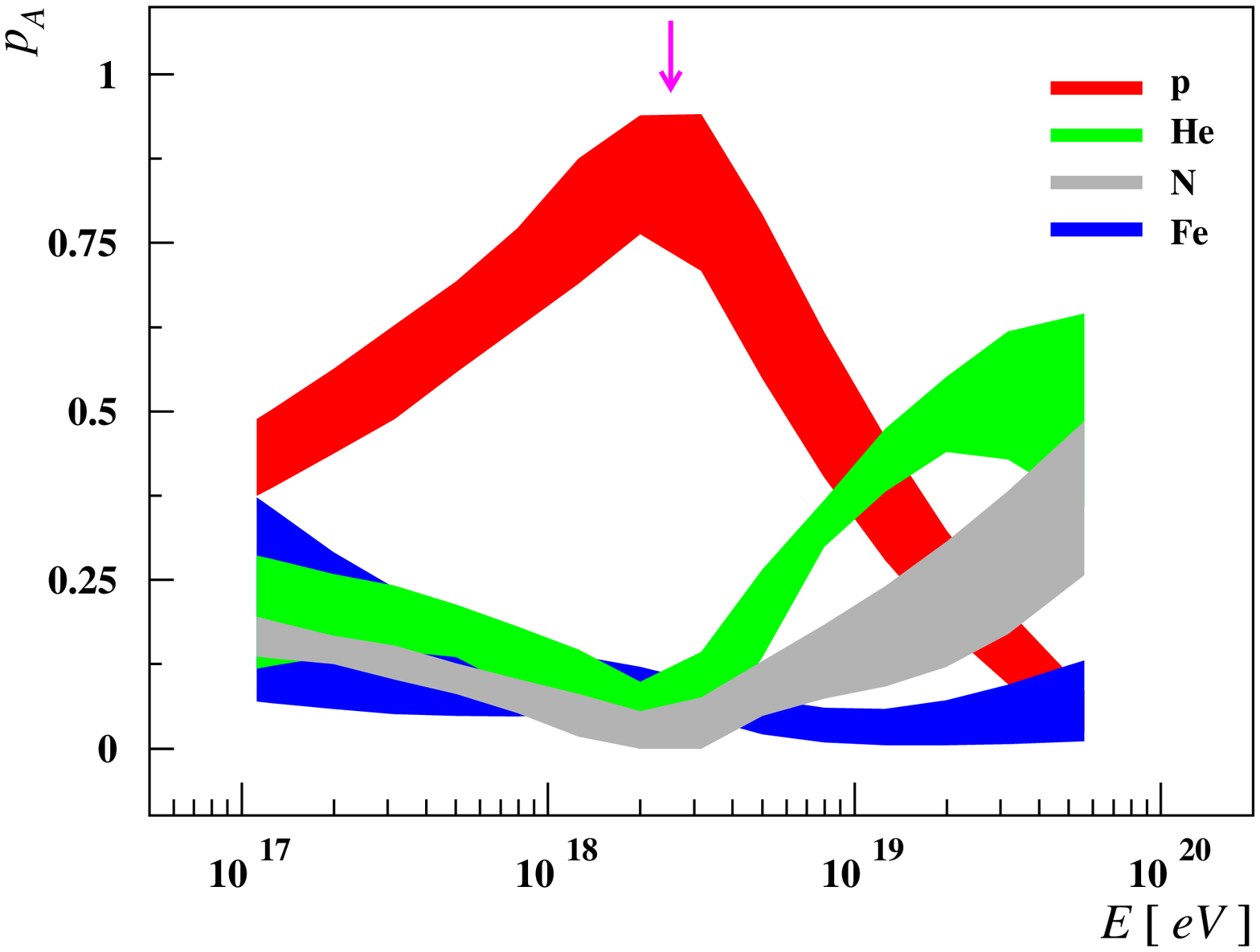}\vspace{-2.16cm}
\includegraphics[width=0.48\textwidth]{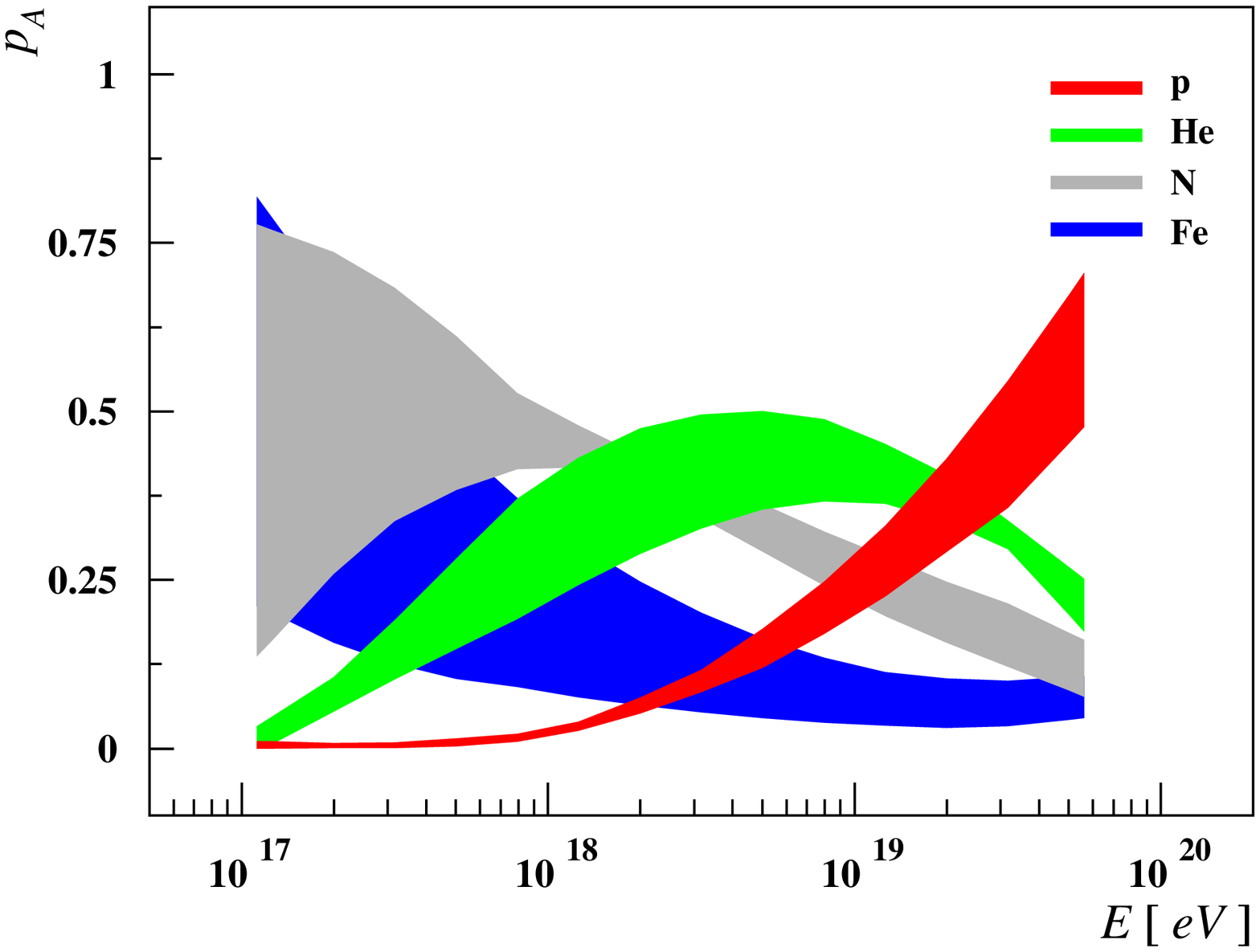}
\caption{
Primary mass partition is depicted as a function of energy.
We used hypothetical shower characteristics for the constant 
elongation rate (top panel) and the elongation rate with a break 
(bottom panel) as shown in~\figs{F01} and~\figg{F02}.
Red, green, gray and blue bands are for proton, helium, nitrogen 
and iron primaries. 
Their widths correspond to uncertainties of parameters 
$C = \Axmax(E_{0})$ and $D$.
}
\label{F03}
\end{figure}

\begin{figure}[t!]
\vspace{-2cm}
\centering
\includegraphics[width=0.48\textwidth]{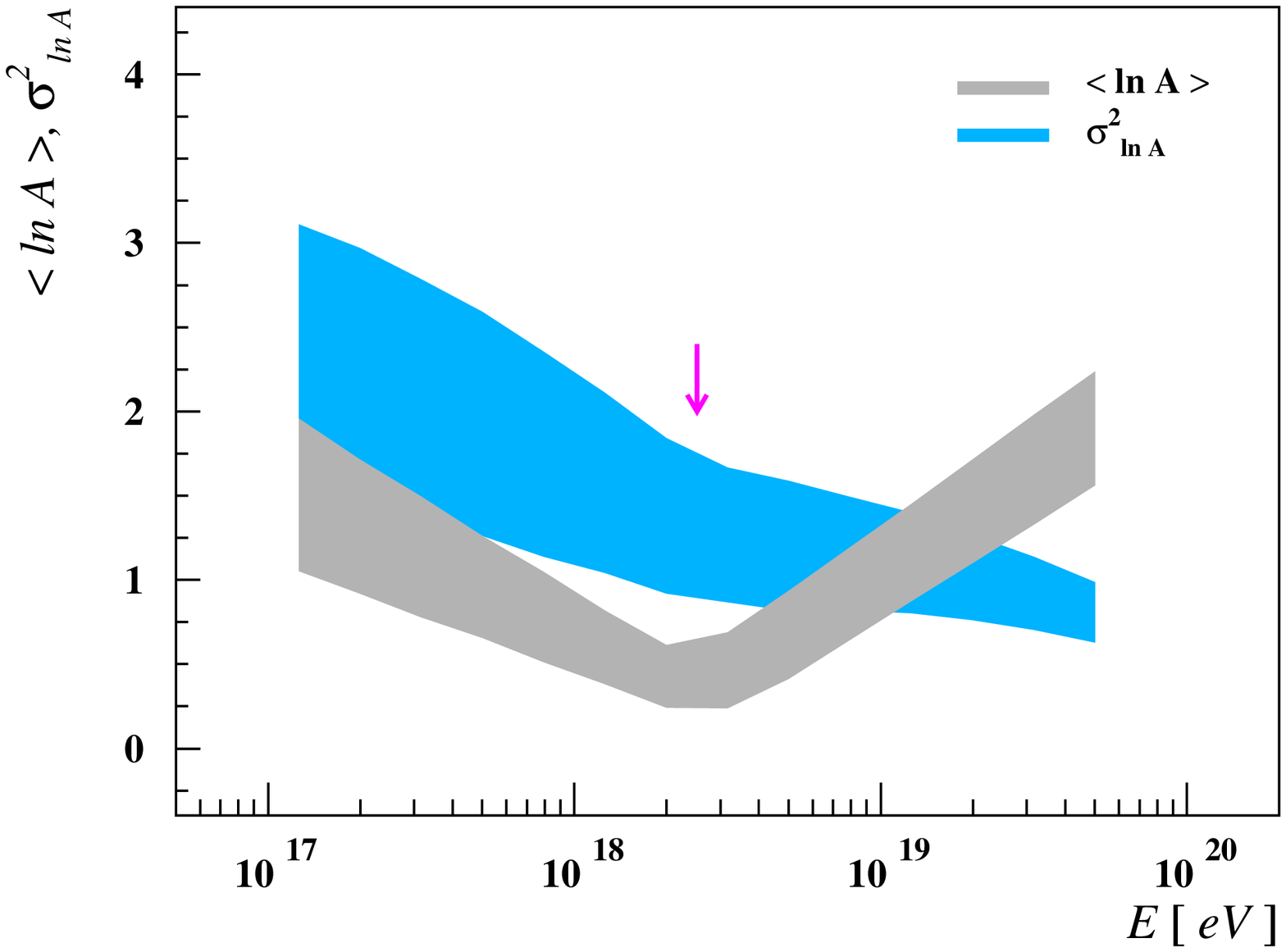}\vspace{-2.16cm}
\includegraphics[width=0.48\textwidth]{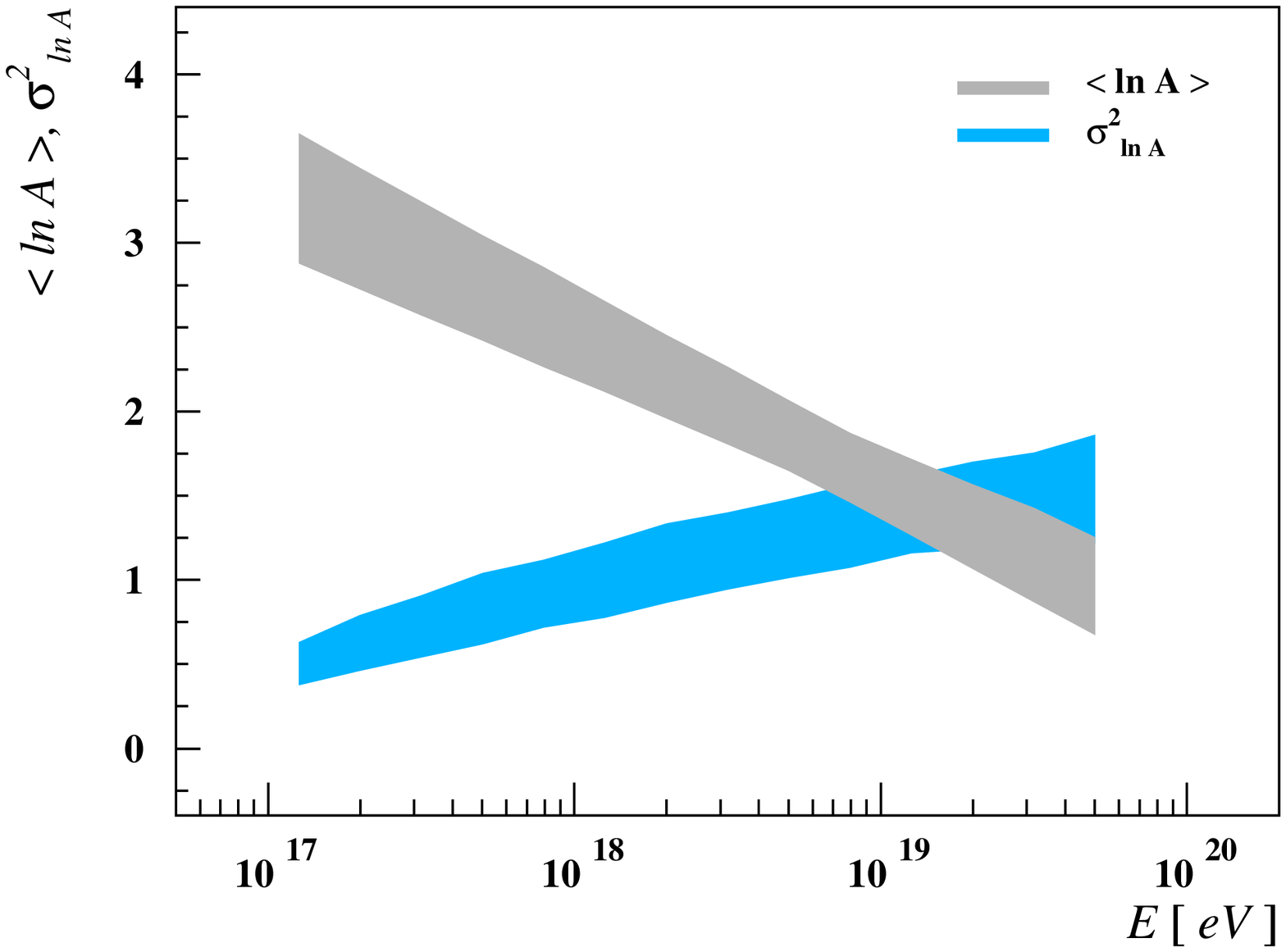}
\caption{
Mean logarithms of the mass number (gray) and their variances 
(blue) are plotted as functions of energy.
Hypothetical shower characteristics for the constant elongation rate 
(top panel) and the elongation rate with a break (bottom panel) 
depicted in~\figs{F01} and~\figg{F02} are used.
Widths of plotted statistics correspond to uncertainties of parameters 
$C = \Axmax(E_{0})$ and $D$.
}
\label{F04}
\end{figure}

Our results are summarized in top panels in~\figs{F03} and~\figg{F04}.
In the top panel in~\fig{F03}, decomposition probabilities of 
hypothetical shower statistics are depicted as functions of energy.
The widths of colored bands correspond to aforementioned uncertainties
in parameters $C$ and $D$.
The mean and variance of logarithmic mass are depicted in the top panel 
in~\fig{F04}.
Both characteristics give the expected trends with steeply falling
$\lnAA$ and growing up variance $\lnAs$ with the increasing energy.
Large uncertainties of heavier primaries at energies where small 
values of $\AXmax$ and $\sxmax$ were chosen are salient features 
of our treatment.

In the second example, we tried to analyze hypothetical shower statistics 
that resemble real data as measured by the Auger detector~\cite{Aug01,Aug02}.
To this end, we prepared the input data, $\AXmax(E)$ and $\sxmax(E)$,
with breaks as depicted in~\figs{F01} and~\figg{F02} by black full points. 
For the energy evolution of the mean depth of shower maximum 
we adopted an elongation rate of $D_{0} = 80~\gIcmS$ for energies 
$\Log(E/\eV) < 18.4$ and $D_{0} = 27~\gIcmS$ above this energy with 
$X_{0} = 708~\gIcmS$, see~\eqa{E07}.
For the evolution of the variance we took  
$s_{0} = 2~\gIcmS$ for $\Log(E/\eV) < 18.4$, $s_{0} = -20~\gIcmS$ 
otherwise, and $\sigma_{0} = 61.2~\gIcmS$.

We adopted the same set of primary species that generated showers 
with statistics under considerations, $p$, ${\rm He}$, ${\rm N}$ 
and ${\rm Fe}$.
We examined a domain for two free parameters giving us
$C = (720-730)~\gIcmS$ and $D = (54-62)~\gIcmS$ ranges where
the best maximum--entropy solutions exist.

The resultant mass decomposition is displayed in the bottom panel
in~\fig{F03}.
The mean value and variance of logarithmic mass are depicted in 
the bottom panel in~\fig{F04}.
Also in this hypothetical example we obtained reasonable solutions.
The chosen breaks in shower statistics are well visible in the energy
evolution of the resultant partition probabilities.
The lightest component, driven up to the chosen break dies out rapidly
after it reaches its maximum value near the break at $\Log(E/\eV) = 18.4$.
Interestingly, the proton--iron mixture is not able to 
explain the chosen energy evolution of shower statistics.
For a reasonable description intermediate mass nuclei are necessary. 

\section{Conclusions}
\label{Sec05}

We used the well justified maximum--entropy method to deduce 
the partition of the mass of CR primaries from the hypothetical 
characteristics of the EAS development that they initiated. 
This method combines simple properties of the generalized Heitler
model, multiplication characteristics of air showers and the measured 
$p$--air cross section. 
It is independent of details on hadronic interactions.

The partition method enables us to establish a reasonable connection 
between the mean value of the logarithmic primary mass number, its 
variance and other observables as well.
The resultant decomposition of the mass distribution describes what 
we know from experiment as effectively as possible provided 
the selected model of the shower evolution holds.
Let us finally stress that the consistency of deduced quantities, 
as interpreted in the Heitler reasoning, is emphasized rather then 
questioned within the partition method. 

\appendix

\section{Partition formalism}
\label{App01}

Let us assume that the quantity $A$ is capable to take $n$ discrete
values $A=1,\dots,n$.
Corresponding probabilities $\pa$ are not known, however. 
Only a set of $r$ expectation values of the functions 
$F_{i}(A)$, $i=1,\dots,r$, $r < n$, is measured.
For setting up a probability distribution which satisfies the given
data, the least biased estimate possible on the basis of partial 
knowledge is used.
This method, known as the maximum--entropy principle, 
is widely used in statistical mechanics~\cite{Jay01}; 
for its statistical background see e.g.~\cite{Rao01}.

Here, Shannon entropy~\cite{Jay01} 
\beql{A01}
S = - k \Ssum{A=1}{n} \pa \ln \pa,
\eeq
where $k$ is a positive constant, is adopted as an information measure 
of the amount of uncertainty in the probability distribution $\pa$ 
of the quantity $A$.
This distribution is determined as the one that maximizes 
entropy $S$ in~\eqa{A01} subject to $r$ constraints, 
$F_{i}(A)$, $i=1,\dots,r$, given their averages that represent
whatever experimental information one has, and subject to 
the normalization condition
\beql{A02}
\langle F_{i} \rangle = \Ssum{A=1}{n} \pa F_{i}(A), \qqa 
i=1,\dots,r, \qqb
\Ssum{A=1}{n} \pa = 1.
\eeq 
Then, the resultant distribution describes what we know about the
quantity $A$ from experiment without assuming anything else~\cite{Jay01}.

In making inferences on the basis of partial information, 
the maximum--entropy probability distribution that maximizes 
Shannon entropy in~\eqa{A01} subject to the experimental 
constraints written in~\eqa{A02} is given by~\cite{Jay01}
\beql{A03}
\pa = Z^{-1}
e^{- \left[ \lambda_{1} F_{1}(A) + \dots + \lambda_{r} F_{r}(A) \right] }, 
\eeq
with the partition function written
\beql{A04}
Z(\lambda_{1}, \dots, \lambda_{r}) = 
\Ssum{A=1}{n} 
e^{- \left[ \lambda_{1} F_{1}(A) + \dots + \lambda_{r} F_{r}(A) \right] },
\eeq
and with Lagrangian multipliers $\lambda_{i}$, $i=1,\dots,r$, 
to be determined. 
The resultant probability distribution obtained in this 
process is spread out as widely as possible without contradicting 
the available experimental information.

\section{Shower variances}
\label{App02}

In our method, the depth of shower maximum caused by a primary proton
with energy $E$ is assumed to be~\cite{Ung01}
\beql{B01}
\Axmax(E) \approx \lambda(E) + X~\ln 
\left( \frac{\kappa E}{2 M \epsilon} \right),
\eeq
where $\lambda(E)$ is the average interaction length for inelastic
$p$--air collisions, $X \approx 37~\gIcmS$ is the radiation
length in air, $\epsilon \approx 84~\MeV$ denotes the critical energy 
in air, $\kappa$ is the elasticity of the first interaction and $M$ 
assigns its multiplicity.
This relationship is well documented by physical arguments and by 
MC simulations as well. 
It can also be derived as an approximate solution of Yule birth process.  

For the variance of the depth of the first interaction we have adopted 
the measured $p$--air cross section at a center of mass energy of 
$\sqrt{s} = 57~\TeV$~\cite{Aug03}.
Relying upon a smooth extrapolation from accelerator measurements, and 
in agreement with model predictions, here we used a parametrization
$\Sigma_{\rm p-Air} \approx \left[ 500 + 50~\Log(E/\EeV) \right]~\mb$.
Within a naive model, the variance of the depth of the first interaction 
is then approximately 
\beql{B02}
\sfr = \sfr(A,E) \approx A^{-\alpha} \xi(E) \sfro, 
\eeq
where $A$ assigns the mass number of a primary CR particle and $\alpha$ 
is a constant index.
The variance of the depth of shower maximum caused by the proton primary
at the reference energy of $E_{0} = 1~\EeV$, $\sfroo \approx 46~\gIcmS$, 
is deduced from the measured $p$--air cross section as well as a function
$\xi(E) \approx 1 - 0.2~\Log(E/\EeV)$.
The $A$--dependent term in~\eqa{B02} accounts for details of 
the first interaction given by individual nucleon--nucleon 
interactions and subsequent nuclear fragmentation~\cite{Ung01}. 
A statistical treatment assuming a subset of interacting nucleons 
supplemented by simple geometrical arguments gives approximately 
$\alpha \approx \frac{2}{3}$.
In our analysis, we have examined values of 
$\alpha \approx 0.3 - 3.0$ yielding slightly different results 
that were negligible if uncertainties of other parameters were 
taken into account.

Assuming an experimental value $\sxmax \approx 60~\gIcmS$
at about $1~\EeV$~\cite{Aug01,Aug02}, and predominantly proton 
primaries, we estimated the variance of the depth of shower 
maximum in the subsequent shower development by
\beql{B03}
\ssh = \ssh(A,E) \approx A^{-1} \ssho,
\eeq
where $\sshoo \approx 38~\gIcmS$. 
The $A$--dependence of the shower variance is given 
by fluctuations in multiplicity $M$ and elasticity $\kappa$ 
of the first (or main) interaction. 
Assuming a model in~\eqa{B01}, the corresponding 
variances caused by primary protons are
$\smuo \approx X^{2} M^{-2} \smu$,
$\skao \approx X^{2} \kappa^{-2} \ska$, 
giving $\ssho = \smuo + \skao$.
In a naive superposition model~\cite{Ung01}, the variance 
of the total multiplicity of $k$ nucleons participating in 
the main interaction with an average multiplicity $\ovl{M}$ 
is $\sigma^{2}(k \ovl{M}) = k^{2} A^{-1} \smu$, and 
similarly for an average elasticity,
$\sigma^{2}(\ovl{\kappa}) = A^{-1} \ska$, 
supporting~\eqa{B03}.

\vspace*{0.5cm}
\footnotesize{{\bf Acknowledgment:}
{This work was supported by the grants MSMT--CR LG13007 and
MSM0021620859 of the Ministry of Education, Youth and Sports 
of the Czech Republic.
The authors would like to thank Petr Travnicek for his help and support.}}


\end{document}